\documentclass[11pt]{article}

\usepackage {graphicx}
\newcommand{\rnc}{\renewcommand}

\def\ds{\dot s}

\def\hA{{\widehat A}}
\def\hB{{\widehat B}}
\def\hC{{\widehat C}}
\def\hD{{\widehat D}}

\makeatletter \@addtoreset{equation}{section}
\rnc{\theequation}{\thesection.\arabic{equation}} \makeatother

\def\sqr#1#2{{\vcenter{\vbox{\hrule height.#2pt
\hbox{\vrule width.#2pt height#1pt \kern#1pt \vrule width.#2pt}\hrule
height.#2pt}}}}
\def\square{\mathchoice\sqr45\sqr45\sqr{2.1}3\sqr{1.5}3}

\def\a{\alpha}

\def\b{\beta}

\def\C{\Gamma}
\def\d{\delta}

\def\D{\Delta}
\def\e{\epsilon}
\def\ve{\varepsilon}

\def\f{\phi}
\def\F{\Phi}
\def\vf{\varphi}
\def\bphi{{\bar\phi}}
\def\h{\eta}
\def\k{\kappa}
\def\l{\lambda}
\def\L{\Lambda}
\def\m{\mu}
\def\n{\nu}
\def\r{\rho}
\def\s{\sigma}
\def\S{\Sigma}
\def\t{\tau}

\def\X{\Xeta}
\def\x{\xi}
\def\O{\Omega}

\def\be{\begin{equation}}
\def\ee{\end{equation}}
\def\bea{\begin{eqnarray}}
\def\eea{\end{eqnarray}}
\def\ba{\begin{array}}
\def\ea{\end{array}}

\def\p{\partial}

\def\ua{\underline{\alpha}}
\def\ub{\underline{\phantom{\alpha}}\!\!\!\beta}

\let\la=\label
\let\bl=\bigl \let\br=\bigr
\let\Br=\Bigr \let\Bl=\Bigl
\let\bm=\bibitem

\def\nn{\nonumber}
\def\bd{\begin{document}}
\def\ed{\end{document}}
\def\ft#1#2{{\textstyle{{\scriptstyle #1}\over {\scriptstyle #2}}}}
\def\fft#1#2{{#1 \over #2}}
\def\sst#1{{\scriptscriptstyle #1}}
\def\oneone{\rlap 1\mkern4mu{\rm l}}
\newcommand{\eq}[1]{(\ref{#1})}
\def\eqs#1#2{(\ref{#1}-\ref{#2})}

\def\det{{\rm det\,}}
\def\tr{{\rm tr}}
\def\Tr{{\rm Tr}}
\def\lra{\leftrightarrow}
\def\fdf{\phi^\dagger\phi}
\def\ffd{\phi\phi^\dagger}
\def\bphi{\overline{\phi}}
\def\qq{\quad\quad}

\newcommand{\w}[1]{\\[0.#1cm]}

\newcommand{\ho}[1]{$\, ^{#1}$}
\newcommand{\hoch}[1]{$\, ^{#1}$}

\def\cramp{\medmuskip = 2mu plus 1mu minus 2mu}
\def\cramper{\medmuskip = 2mu plus 1mu minus 2mu}
\def\crampest{\medmuskip = 1mu plus 1mu minus 1mu}
\def\uncramp{\medmuskip = 4mu plus 2mu minus 4mu}
\def\ben{\begin{equation}}
\def\een{\end{equation}}
\def\half{{\textstyle{1\over2}}}
\let\a=\alpha \let\b=\beta \let\g=\gamma \let\d=\delta \let\e=\epsilon
\let\z=\zeta \let\h=\eta \let\q=\theta \let\i=\iota \let\k=\kappa
\let\l=\lambda \let\m=\mu \let\n=\nu \let\x=\xi
\let\r=\rho
\let\s=\sigma \let\t=\tau \let\f=\phi  \let\y=\psi
\let\D=\Delta \let\Q=\Theta \let\L=\Lambda
\let\X=\Xi \let\P=\Pi \let\S=\Sigma \let\U=\Upsilon \let\F=\Phi \let\Y=\Psi
\let\W=\Omega
\let\la=\label \let\ci=\cite \let\re=\ref
\let\se=\section \let\sse=\subsection \let\ssse=\subsubsection
\def\nn{\nonumber} \def\bd{\begin{document}} \def\ed{\end{document}}
\def\ds{\documentstyle} \let\fr=\frac \let\bl=\bigl \let\br=\bigr
\let\Br=\Bigr \let\Bl=\Bigl
\let\bm=\bibitem
\let\na=\nabla
\let\pa=\partial \let\ov=\overline
\def\ba{\begin{array}}
\def\ea{\end{array}}
\def\ft#1#2{{\textstyle{{\scriptstyle #1}\over {\scriptstyle #2}}}}
\def\fft#1#2{{#1 \over #2}}
\def\del{\partial}
\def\vp{\varphi}
\def\sst#1{{\scriptscriptstyle #1}}
\def\oneone{\rlap 1\mkern4mu{\rm l}}
\def\td{\tilde}
\def\ie{\rm i.e.\ }
\def\dalemb#1#2{{\vbox{\hrule height .#2pt
        \hbox{\vrule width.#2pt height#1pt \kern#1pt
                \vrule width.#2pt}
        \hrule height.#2pt}}}
\def\square{\mathord{\dalemb{6.8}{7}\hbox{\hskip1pt}}}
\newcommand{\ap}{\alpha^\prime}
\newcommand{\bp}{\tilde \beta^\prime}
\def\0{{\sst{(0)}}}
\def\1{{\sst{(1)}}}
\def\2{{\sst{(2)}}}
\def\3{{\sst{(3)}}}
\def\4{{\sst{(4)}}}
\def\5{{\sst{(5)}}}
\def\6{{\sst{(6)}}}
\def\7{{\sst{(7)}}}
\def\8{{\sst{(8)}}}
\def\n{{\sst{(n)}}}
\def\cA{{{\cal A}}}
\def\cF{{{\cal F}}}
\def\tV{\widetilde V}
\def\tW{\widetilde W}
\def\tH{\widetilde H}
\def\tE{\widetilde E}
\def\tF{\widetilde F}
\def\tA{\widetilde A}
\def\im{{{\rm i}}}
\def\tY{{{\wtd Y}}}
\def\ep{{\epsilon}}
\def\vep{{\varepsilon}}
\def\R{\rlap{\rm I}\mkern3mu{\rm R}}
\def\bD{{{\bar D}}}

\def\R{\rlap{\rm I}\mkern3mu{\rm R}}
\def\bD{{{\bar D}}}
\def\R{{{\Bbb R}}}
\def\CN{{{\Bbb C}}}
\def\H{{{\Bbb H}}}
\def\CP{{{\Bbb C}{\Bbb P}}}
\def\RP{{{\Bbb R}{\Bbb P}}}
\def\Z{{{\Bbb Z}}}
\def\bA{{{\Bbb A}}}
\def\bB{{{\Bbb B}}}
\def\bC{{{\Bbb C}}}
\def\bR{{{\Bbb R}}}
\def\bD{{{\Bbb D}}}
\def\bE{{{\Bbb E}}}
\def\bZ{{{\Bbb Z}}}
\def\Re{{{\frak{Re}}}}
\def\Im{{{\frak{Im}}}}
\def\cosec{{\,\hbox{cosec}\,}}
\def\Gm{{\Gamma_{\!\! -}}}
\def\Gp{{\Gamma_{\!\! +}}}
\def\stan{{standard }}
\def\nonstan{{supernumerary }}

\def\cosech{{\hbox{cosech}}}

\def\etcyc{{\hbox{and cyclic}}}
\def\btheta{{\bar\theta}}

\newcommand{\tamphys}{\it\small George P. and Cynthia W. Mitchell Institute for
Fundamental Physics\\
Texas A\&M University, College Station, TX 77843-4242}

\newcommand{\boun}{\it\small Bo\u{g}azi\c{c}i University,
Department of Physics, 34342, Bebek, Istanbul, Turkey}

\newcommand{\feza}{\it\small Feza G\"ursey Institute, 34684, \c{C}engelk\"{o}y,
Istanbul, Turkey}

\newcommand{\auth}{\large  D.C. Jong\hoch{1}, A. Kaya\hoch{2,3} and
E. Sezgin\hoch{1,2,3}}

\begin{document}

\hfill{MIFP-06-23}

\hfill{hep-th/0608034}

\vspace{20pt}

\begin{center}


{\Large\bf $6D$ Dyonic String With Active Hyperscalars }


\vspace{30pt}

\auth

\vspace{15pt}

\begin{itemize}
\item [$^1$]\tamphys
\item [$^2$]\boun
\item[$^3$] \feza
\end{itemize}

\vspace{60pt}

{\sc Abstract}

\end{center}

We derive the necessary and sufficient conditions for the existence
of a Killing spinor in $N=(1,0)$ gauge supergravity in six
dimensions coupled to a single tensor multiplet, vector multiplets
and hypermultiplets. These are shown to imply most of the field
equations and the remaining ones are determined. In this framework,
we find a novel $1/8$ supersymmetric dyonic string solution with
nonvanishing hypermultiplet scalars. The activated scalars
parametrize a $4$ dimensional submanifold of a quaternionic
hyperbolic ball. We employ an identity map between this submanifold
and the internal space transverse to the string worldsheet. The
internal space forms  a $4$ dimensional analog of the
Gell-Mann-Zwiebach tear-drop which is noncompact with finite volume.
While the electric charge carried by the dyonic string is arbitrary,
the magnetic charge is fixed in Planckian units, and hence
necessarily non-vanishing.  The source term needed to balance a
delta function type singularity at the origin is determined. The
solution is also shown to have $1/4$ supersymmetric  $AdS_3\times
S^3$ near horizon limit where the radii are proportional to the
electric charge.

\vskip 10pt

\pagebreak

\tableofcontents

\newpage


\section{Introduction}


Anomaly-free matter coupled supergravities in six dimensions
naturally arise in $K3$ compactification of Type I and heterotic
string theories \cite{Green:1984bx}. Owing to the fact that $K3$ has
no isometries, all of the resulting $6D$ models are ungauged in the
sense that the $R$-symmetry group $Sp(1)_R$, or its $U(1)_R$
subgroup thereof, is only a global symmetry. The $R$-symmetry gauged
general matter coupled models, on the other hand, have been
constructed directly in six dimensions long ago
\cite{Nishino:1986dc,Nishino:1997ff}. These theories harbor
gravitational, gauge and mixed anomalies which can be encoded in an
$8$-form anomaly polynomial, and the Green-Schwarz anomaly
cancelation mechanism requires its factorization. It turns out that
the $R$-symmetry gauging reduces drastically the space of solutions
to this requirement.

At present, the only known  ``naturally'' anomaly-free gauged
supergravities in $6D$ are:

\begin{itemize}

\item the $E_7\times E_6 \times U(1)_R$ invariant model in which
the hyperfermions are in the $(912,1,1)$ representation of the gauge
group \cite{Randjbar-Daemi:1985wc},

\item the $E_7\times G_2\times U(1)_R$ invariant model with
hyperfermions in the $(56,14,1)$ representation of the gauge group
\cite{Avramis:2005qt}, and

\item the $F_4\times Sp(9)\times U(1)_R$ invariant model with
hyperfermions in the $(52,18,1)$ representation of the gauge group
\cite{Avramis:2005hc}.

\end{itemize}

The anomaly freedom of these models is highly nontrivial, and they
are natural in the sense that they do not contain any gauge-singlet
hyperfermions. If one considers a large factor of $U(1)$ groups, and
tune their $U(1)$ charges in a rather ad-hoc way
\cite{Avramis:2005hc}, or considers only products of $SU(2)$ and
$U(1)$ factors with a large number of hyperfermions, and tune their
$U(1)$ charges again in an ad-hoc way, infinitely many possible
anomaly-free combinations arise \cite{Suzuki:2005vu}. These models
appear to be ``unnatural'' at this time.

In fact, none of the above mentioned models, natural or not, have
any known string/M-theory origin so far, though progress has been
made in embedding \cite{Cvetic:2003xr} a minimal sub-sector with
$U(1)_R$ symmetry and no hyperfermions \cite{Salam:1984cj} in
string/M theory. An apparently inconclusive effort has also been
made in \cite{Avramis:2004cn} in which the $6D$ theory is considered
to live on the boundary of a $7D$ theory, which, in turn is to be
obtained from string/M-theory.

Finding the string/M-theory origin of the anomaly free models
mentioned above is likely to uncover some interesting mechanisms for
descending to lower dimensions starting from string/M-theory.
Moreover, models of these type have been increasingly finding
remarkable applications in cosmology and braneworld scenarios
\cite{Halliwell:1986bs,Maeda:1985es,Aghababaie:2003wz,Gibbons:2003di,Nair:2004yu,Carter:2006uk}.

In this paper, we will not address the string/M-theory origin of the
$6D$ theories at hand but rather investigate the general form of
their supersymmetric solutions, and present, in particular, a dyonic
string solution in which the hyperscalar fields have been activated.
Our aims are:

\begin{itemize}

\item to lay out the framework for finding further solutions which,
in turn, may lead to new solutions in other theories of interest
that live in diverse dimensions,

\item to establish the fact that (dyonic) string solution exists
in a more general situation than so far that has been known, in the
sense that new type of fields, to wit, hyperscalars, have been
activated, and

\item  to open new avenues in the compactification schemes in which
the sigma model sector of supergravity theories are exploited.

\end{itemize}

These aims call for a modest summary of what has been done in these
areas so far. To begin with, the general form of supersymmetric
solutions in $6D$ have been studied in
\cite{Gutowski:2003rg,Cariglia:2004kk}, though in the absence of
hypermultiplets. We will fill this gap here. We will extend the
analysis for the existence of Killing spinors, determine the
resulting integrability conditions and the necessary and sufficient
equations for finding exact solutions, without having to directly
solve all the field equations.

Second, various dyonic string solutions of $6D$ supergravities exist
in the literature
\cite{Duff:1995yh,Duff:1996cf,Guven:2003uw,Randjbar-Daemi:2004qr},
though again, none of them employ the hypermatter. We will find some
novel features here such as the necessity to switch on the magnetic
charge of the dyonic string.

Third, concerning the use of (higher than one dimensional) sigma
model sector of supergravity theories in finding exact solutions, in
the case of ungauged supergravities  the oldest result is due to
Gell-Mann-Zwiebach \cite{Gell-Mann:1984mu} who found the
half-supersymmetry breaking tear-drop solution of Type IIB
supergravity, by exploiting its $SU(1,1)/U(1)$ sigma model sector.
The tear-drop represents the two-dimensional internal space which is
non-compact with finite volume. The sigma model sector of Type IIB
supergravity has also been utilized in finding an instanton solution
dual to a $7$-brane \cite{Gibbons:1995vg}. Supersymmetric two
dimensional tear-drop solutions in ungauged $D<10$ supergravities
are also known
\cite{Gell-Mann:1984mu,Izquierdo:1994jz,Gibbons:2003di,Nair:2004yu,Kehagias:2005dp}.
More recently, the general form of the supersymmetric solutions in
ungauged $4D$ supergravities, including their coupling to
hypermatter, have been provided in \cite{Hubscher:2006mr}.

In the case gauged supergravities, a solution of the matter coupled
$N=(1,0)$ gauged supergravity in $6D$ called 'the superswirl' has
been found in \cite{Parameswaran:2005mm} where two hyperscalars are
activated. One of these scalars is dilatonic and the other one is
axionic. Supersymmetric domain-wall solutions of maximal gauged
supergravities in diverse dimensions where only the dilatonic
scalars of the sigma model are activated have appeared in
\cite{Bergshoeff:2004nq}. Supersymmetric black string solutions of
matter coupled $N=2, D=3$ gauged supergravity exists in which only a
single dilaton is activated in the Kahler sigma model sector
\cite{Deger:1999st}. In such models, supersymmetric solutions
 with the additional axionic scalars activated, have also been found
\cite{Abou-Zeid:2001tu,Deger:2004mw,Deger:2006uc}. Finally,
conditions for Killing spinors and general form of the
supersymmetric solutions in matter coupled gauged supergravities in
$N=2, D=5$ supergravities have also been investigated
\cite{Cacciatori:2002qx} but no specific solutions with
multi-hyperscalars activated seem to have appeared.

To summarize, we see that there exist only few scattered results on
the nontrivial use of gauged sigma models in supergravity theories
in finding exact supersymmetric solutions. As stated earlier, one of
our goals in this paper is to take a step towards a systematic
approach to this problem.  We shall come back to this point in the
Conclusions.

Turning to the tear-drop solutions, a key feature in these
backgrounds is the identity map by which the scalars of the sigma
model manifold are identified with those of the internal part of the
spacetime. The brief summary of literature above only dealt with
solutions that have supersymmetry. The idea of identity map, on the
other hand, was first proposed by Omero and Percacci
\cite{Omero:1980vx} long ago in the context of bosonic sigma models
coupled to gravity. This work was generalized later in
\cite{Ianus:1987xa}. Several more papers may well exist that deal
with the solutions of sigma model coupled ordinary gravities, as
opposed to supergravities, but we shall not attempt to survey them
since our emphasis is on gauge supergravities with sigma model
sectors in this paper.

After the description of the matter coupled $6D$ supergravity in the
next section, the conditions for the existence of Killing spinors,
and their integrability conditions will be presented in Sections 3
and 4, respectively. The new dyonic string solution and its
properties are then described in Sections 5 and 6, respectively. The
summary of our results that emphasizes the key points, and selected
open problems are given in the Conclusions. Three appendices that
contain our conventions and useful formulae are also presented.


\section{The Model}


\subsection{ Field Content and the Quaternionic Kahler Scalar Manifold}


\bigskip

The six-dimensional gauged supergravity model we shall study
involves the combined $N=(1,0)$ supergravity plus anti-selfdual
supermultiplet $(g_{\mu\nu}, B_{\mu\nu}, \varphi , \psi_{\mu +}^A,
\chi_{-}^A)$, Yang-Mills multiplet $(A_\mu, \lambda_{+}^A)$ and
hypermultiplet $(\phi^\a, \psi_{-}^a)$. All the spinors are
symplectic Majorana-Weyl, $A=1,2$ label the doublet of the $R$
symmetry group $Sp(1)_R$ and $a=1,...,2n_H$ labels the fundamental
representation of $Sp(n_H)$. The chiralities of the fermions are
denoted by $\pm$.

The hyperscalars $\phi^\a,\ \a=1,..., 4n_H$ parameterize the coset
$~Sp(n_H,1)/Sp(n_H)\otimes Sp(1)_R$. This choice is due to its
notational simplicity. Our formulae can straightforwardly be adapted
to more general quaternionic coset spaces $G/H$ whose list can be
found, for example in \cite{Bagger:1983tt}. In this paper, we gauge
the group
\be K\times Sp(1)_R \subset Sp(n_H,1)\ ,\quad\quad K\subseteq
Sp(n_H)\ .\nn\ee

The group $K$ is taken to be semi-simple, and the $Sp(1)_R$ part of
the gauge group can easily be replaced by its $U(1)_R$ subgroup.

We proceed by defining the basic building blocks of the model
constructed in \cite{Nishino:1986dc} in an alternative notation. The
vielbein $V_\a^{aA}$, the $Sp(n_H)$ composite connection $Q_\a^{ab}$
and the $Sp(1)_R$ composite connection $Q_\a^{AB}$ on the coset are
defined via the Maurer-Cartan form as
\be L^{-1} \p_\a L= V_\a^{aA}T_{aA} + \ft12\, Q_\a^{ab}T_{ab}+
\ft12\,Q_\a^{AB}T_{AB}\ , \la{mc1} \ee
where $L$ is the coset representative, $(T_{ab},T_{AB}, iT_{aA})\equiv T_{\hA\hB}\ $
obey the $Sp(n_H,1)$ algebra
\bea
&&[T_{\hA\hB},T_{\hC\hD}]= -\Omega_{\hB\hC} T_{\hA\hD}
-\Omega_{\hA\hC} T_{\hB\hD} -\Omega_{\hB\hD} T_{\hA\hC}
-\Omega_{\hA\hD} T_{\hB\hC} \ ,\nn\w3
&&\Omega_{\hA\hB}= \left(
                      \begin{array}{cc}
                        \epsilon_{AB} & 0 \\
                        0 & \Omega_{ab} \\
                      \end{array}
                    \right)
\la{alg} \, .\eea
The generator $T_{aA}$ is hermitian and $(T_{AB}, T_{ab})$ are
anti-hermitian. The vielbeins obey the following relations:
\be g_{\a\b}V^\a_{a A} V^\b_{bB}=\Omega_{ab}\epsilon_{AB}\ ,\quad\quad
V^\a_{aA} V^{\b aB} +\a \lra \b = g^{\a\b} \delta_A^B\ , \label{vg}\ee
where $g_{\a\b}$ is the metric on the coset. Another useful
definition is that of the three quaternionic Kahler structures given
by
\be V^A_{\a a} V_\b^{aB} - A\lra B = 2J^{AB}_{\a\b} \ . \la{j}\ee
Next, we define the components of the gauged Maurer-Cartan form as
\be L^{-1} D_\mu L = P_\mu^{aA}T_{aA} + \ft12\, Q_\mu^{ab}T_{ab}+
\ft12\,Q_\mu^{AB}T_{AB}    \ ,\la{mc2} \ee
where
\be D_\mu L=\left( \partial_\mu - A_\mu^I T^I  \right) L\
,\la{mc3}\ee
$A_\mu^I$ are the gauge fields of $K \times Sp(1)_R$. All gauge
coupling constants are set equal to unity for simplicity in
notation. They can straightforwardly be re-instated. We also use
the notation
\be T^I= (T^{I'}, T^r)\ ,\qquad T_r= 2 T_r^{AB}\,T_{AB}\ ,\ \qquad
T^r_{AB}= -\ft{i}2\,\sigma^r_{AB}\ , \quad  r=1,2,3\ .\ee
The components of the Maurer-Cartan form can be expressed in terms
of the covariant derivative of the scalar fields as follows
\cite{Percacci:1998ag}
\be P_\m^{aA}=  (D_\m\phi^\a ) V_\a^{aA} \ ,\quad\quad
Q_\m^{ab}=(D_\m\phi^\a ) Q_\a^{ab}-A_\mu^{ab}\ ,\quad\quad
Q_\m^{AB}=(D_\m\phi^\a) Q_\a^{AB}-A_\m^{AB}\ ,\la{df}\ee
where
\be D_\m \phi^\a = \p_\m \phi^\a -A_\m^I K^{I\a}\ ,\la{df2}\ee
and $K^I(\phi)$ are the Killing vectors that generate the $K\times
Sp(1)_R$ transformations on $G/H$.

Other building blocks to define the model are certain $C$-functions
on the coset. These were defined in \cite{Nishino:1997ff}, and
studied further in \cite{Percacci:1998ag} where it was shown that
they can be expressed as
\be L^{-1} T^I L \equiv C^I = C^{IaA}T_{aA}+\ft12
C^{IAB}T_{AB}+\ft12C^{Iab}T_{ab}\, . \ee
Differentiating and using the algebra \eq{alg} gives the useful
relation
\be D_\mu C^I = \left(P_\mu^a{}_B C^{IAB}+P_{\mu
b}{}^AC^{Iab}\right)\,T_{aA} + P_\mu^{aA}C^I_a{}^B\, T_{AB}
+P_\mu^{aA}C^{Ib}{}_A\,T_{ab}\ . \la{dc}\ee
Moreover, using \eq{mc2} and \eq{df} we learn that
\be K^{I\a}V_\a^{aA} = C^{IaA}\ ,\qquad K^{I\a}Q_\a^{ab}=
C^{Iab}-\d^{II'}T_{I'}^{ab}\ ,
\qquad K^{I\a} Q_\a^{AB}= C^{IAB}-\d^{Ir}\,T_r^{AB}\ .\la{kv}\\
\ee
Finally, it is straightforward and useful to derive the identities
\bea D_{[\mu} P_{\nu]}^{aA}  &=& -\ft12\, F_{\mu\nu}^I C^{IaA}\
,\la{id1}\w2
P_{[\mu}^{aA}  P_{\nu]}^b{}_A &=& \ft12\, Q_{\mu\nu}^{ab} +\ft12
F_{\mu\nu}^I C^{Iab}\ ,\la{id2}\w2
P_{[\mu}^{aA} P_{\nu] a}{}^B &=& \ft12\,Q_{\mu\nu}^{AB} +\ft12
F_{\mu\nu}^I C^{IAB}\ . \la{id3}\eea


\subsection{ Field Equations and Supersymmetry Transformation Rules}


\bigskip

The Lagrangian for the anomaly free model we are studying can be
obtained from \cite{Nishino:1986dc} or \cite{Nishino:1997ff}. We
shall use the latter in the absence of Lorentz Chern-Simons terms
and Green-Schwarz anomaly counterterms. Thus, the bosonic sector of
the Lagrangian is given by \cite{Nishino:1997ff}
\be
e^{-1}{\cal L} =R\,- \ft14 (\p\vf)^2- \ft1{12} e^\vf\,
G_{\mu\nu\rho}G^{\mu\nu\rho}- \ft14\,e^{\ft12\vf}\,
F^I_{\m\nu}\,F^{I\m\nu}\,
 -2P_\m^{aA}\,P^\m_{aA}- 4 \,
e^{-\ft12\vf}\,C^I_{AB} C^{IAB}\ ,\ee
where the Yang-Mills field strength is defined by $F^I=dA^I +\ft12
f^{IJK} A^J\wedge A^K$ and $G$ obeys the Bianchi identity
\be dG = \ft12 F^I\wedge F^I\ .\nn\w2 \label{bianchi}\ee
The bosonic field equations following from the above Lagrangian are
\cite{Nishino:1997ff}
\bea
R_{\m\nu} &=& \ft14 \del_\m\vf\, \del_\nu\vf + \ft12 e^{\ft12\vf}\,
(F^2_{\m\nu} - \ft18 F^2\,g_{\m\nu}) + \ft14 e^\vf\, (G^2_{\m\nu} -
\ft16 G^2\, g_{\m\nu}) \nn\w2 
&& -2 P_\m^{aA} P_{\nu aA} + e^{-\ft12\vf}(C^I_{AB}C^{IAB})  \ g_{\m\nu}\
, \nn\w2
\square\, \vf&=& \ft14 e^{\ft12\vf}\,F^2 + \ft16 e^\vf\, G^2 -4\,
e^{-\ft12\vf}\,C^I_{AB} C^{IAB} \nn\w2 
D_\rho\big(e^{\ft12\vf}\,F^{I\rho}{}_{\m}\big) &=& \ft12e^\vf\,
F^{I\rho\s}G_{\rho\s\m} +4 P^{aA}_\m C^I_{aA}\ ,\nn\w2 
\nabla_\rho\left(e^\vf\, G^{\rho}{}_{\m\nu}\right) &=& 0\ ,\nn\w2 
D_\m P^{\m aA}  &=& 4 e^{-\ft12\vf} C^{IAB} C^{Ia}{}_{B}\ , \la{e7}
\eea
where we have used a notation
$V^2_{\mu\nu}=V_{\mu\lambda_2...\lambda_p}
V_\nu{}^{\lambda_2...\lambda_p}$ and $V^2=g^{\mu\nu}V_{\mu\nu}$ for
a $p$-form $V$, and $F^2=F_{\mu\nu}^I F^{\mu\nu I}$. The local
supersymmetry transformations of the fermions, up to cubic fermion
terms that will not effect our results for the Killing spinors, are
given by \cite{Nishino:1997ff}
\bea
\d \psi_\m &=& D_\m \ve + \ft1{48} e^{\ft12\vf}
G_{\nu\s\rho}^+\,\C^{\nu\s\rho}\, \C_\m\,\ve \ ,\la{s1}\w2
\d\chi &=&\ft14\left( \C^\m\p_\m \vf  -\ft16 e^{\ft12\vf}
G_{\m\nu\rho}^-\,\C^{\m\nu\rho} \right)\ve\ , \la{s2}\w2
\d \l_A^I &=& -\ft18 F_{\m\nu}^I\C^{\m\nu}\ve_A
    -e^{-\ft12\vf} C^I_{AB}~\ve^B \ ,
  \la{s3}\w2
\d\psi^a &=&  P_\m^{a A} \C^\m \ve_A \ , \la{s4}
\eea

where $D_\m\ve_A = \nabla_\m\ve_A + Q_{\m A}{}^B \ve_B$, with
$\nabla_\mu$ containing the standard torsion-free Lorentz connection
only. The transformation rules for the gauge fermions differ from
those in \cite{Nishino:1986dc} by a field redefinition.


\section{ Killing Spinor Conditions}


The Killing spinor in the present context is defined to be the
spinor of the supersymmetry transformations which satisfies the
vanishing of the supersymmetric variations of all the spinors in the
model. The well known advantage of seeking such spinors is that the
necessary and sufficient conditions for their existence are first
order equations which are much easier than the second order field
equations, and moreover, once they are solved, the integrability
conditions for their existence can be shown to imply most of the
field equations automatically. In deriving the necessary and
sufficient conditions for the existence of Killing spinors, it is
convenient to begin with the construction of the nonvanishing
fermionic bilinears, which provide a convenient tool for analyzing
these conditions. In this section, firstly the construction and
analysis of the fermionic bilinears are given, and then all the
necessary and sufficient conditions for the existence of Killing
spinor are derived.


\subsection{Fermionic Bilinears and Their Algebraic Properties}


There are only two nonvanishing fermionic bilinears that can be
constructed from {\it commuting}  symplectic-Majorana spinor $\e^A$.
These are:
\bea  {\bar\e}^A\C_\mu\e^B &\equiv&  V_\mu^{AB}\ ,\nn\\
{\bar\e}^A\C_{\mu\nu\rho}\e^B &\equiv& X^r_{\mu\nu\rho} T_r^{AB}\ .
\eea
Note that $X^r$ is a self-dual three-form due to chirality
properties. From the Fierz identity
${\C_\mu}_{(\a\b}\C^\mu_{\gamma)\delta}=0$, it follows that
\bea V^\mu V_\mu=0\ ,\qquad  i_V X^r =0\ . \eea
Introducing the orthonormal basis
\be ds^2=2e^+e^- + e^ie^i\ , \ee
and identifying
\be e^+=V\ , \ee
the equation $i_V X^r=0$ and self-duality of $X^r$ yield
\be X^r= 2 V\wedge I^r\ , \la{xv}\ee
where
\be I^r=\ft12 I^r_{ij}\,e^i\wedge e^j\  \ee
is anti-self dual in the 4-dimensional metric $ds_4^2=e^ie^i$.
Straightforward manipulations involving Fierz identities imply that
$I^r$ are quaternionic structures obeying the defining relation
\be (I^r)^i{}_k\,(I^s)^k{}_j = \e^{rst} (I^t)^i{}_j -
\delta^{rs}\delta^i_j\ .\label{hks} \ee
Finally, using the Fierz identity ${\C_\mu}_{(\a\b}\C^\mu_{\gamma)\delta}=0$
once more, one finds that
\be V_\mu\C^\mu \e=\C^+\e=0\ .\la{cplus}  \label{susy1}\ee
If there exists more than one linearly independent Killing spinor,
one can construct as many linearly independent null vectors. In this
case \eq{cplus} is obeyed by each Killing spinor and the
corresponding null vector, i.e. $V_\mu^1\Gamma^\mu \e_1=0,\
V_\mu^2\Gamma^\mu \e_2=0$, but it may be that $V_\mu^1\Gamma^\mu
\e_2 \ne 0$ and/or $V_\mu^2\Gamma^\mu \e_1\ne 0$. In that case,
\eq{cplus} should be relaxed since $\e$ should be considered as a
linear combination of $\e_1$ and $\e_2$.

\subsection{Conditions From $\delta\lambda^I=0$}


Multiplying  \eq{s3} with ${\bar\e}^B \Gamma^\rho$, we obtain
\bea i_V F^I &=&0\ ,\la{f1}\\
F^{Iij} I^{r}_{ij}&=&4 e^{-\ft12\vf}\,C^{Ir}\ .\la{f2} \eea
The second has been simplified by making use of \eq{f1} and \eq{xv}.
Multiplying \eq{s3} with ${\bar\e}^B\C_{\lambda\tau\rho}$, on the
other hand, gives
\bea && F^I\wedge V + \star (F^I\wedge V)+2 e^{\ft12\vf}\,C^{Ir}
X^r=0\ , \la{f3}\\
&& \ft34 F^{I\sigma}{}_{[\mu}X^{r}_{\nu\rho]\sigma}+\ft12\e^{rst}
e^{-\ft12 \vf}C^{Is}X^{t}_{\mu\nu\rho}=0\ .\la{f4}\eea
One can show that these two equations are identically satisfied upon the
use of \eq{f1} and \eq{f2}, which, in turn imply that $F$ must take
the form
\be F^I=-e^{-\ft12\vf}\,C^{Ir} I^r+{\widetilde F}^I+V\wedge
\omega^I\ ,\la{fi} \ee
where ${\widetilde F}^I=\ft12 {\widetilde F}^I_{ij}\,e^i\wedge e^j$
is self-dual, and $\omega^I=\omega^I_i\, e^i$. Reinstating the gauge
coupling constants, we note that the $C$-function dependent term
will be absent when the index $I$ points in the direction of a
subgroup of $K \subset Sp(2n_H)$ under which all the hyperscalars
are neutral.

Substituting \eq{fi} into the supersymmetry transformation rule, and
recalling \eq{cplus}, one finds that  \eq{s3} gives the additional
conditions on the Killing spinor
\be \left(\ft18 I^r_{ij}\C^{ij} \delta^A_B
-T^{rA}{}_B\right)\,\e^B=0\ . \la{it}\ee
The contribution from ${\widetilde F}$ drops out due to
chirality-duality properties involved. Writing this equation as
${\cal O}^r \e=0$, one can check that $[{\cal O}^r,{\cal O}^s]
=\e^{rst}{\cal O}^t$. Thus, any two projection imply the third one.

In summary, the necessary and sufficient conditions for
$\delta\lambda^I=0$ are \eq{fi} and \eq{it}.


\subsection{Conditions From $\delta\psi^a=0$}


This time  multiplying  \eq{s3} with ${\bar\e}^B$ and
${\bar\e}^B\C_{\lambda\tau}$ gives rise to four equations which can be shown to imply
\bea V^\mu P_\mu^{aA}&=& 0\ ,\la{h1}\w2 P_i^{aA} &=&
2(I^r)_i{}^j\,(T^r)^A{}_B\,P_j^{aB}\ . \la{h2} \eea
Using \eq{j} and \eq{df}, we can equivalently reexpress the second
equation above as
\be D_i\phi^\a = (I^r)_i{}^j\,(J^r)_\b{}^\a\, D_j\phi^\b\ . \la{hc}
\ee
Writing \eq{h2} as $P^a={\cal O} P^a$, we find that $({\cal
O}-1)({\cal O}-3 )=0$. Thus, \eq{h2} implies that $P^a$ is an
eigenvector of ${\cal O}$ with eigenvalue one. Moreover, using
\eq{h2} directly in the supersymmetry transformation rule \eq{s4},
and using the projection condition \eq{it}, we find that
$\delta\psi^a=3\delta\psi^a$, and hence vanishing.

In summary, the necessary and sufficient conditions for
$\delta\psi^a=0$ are \eq{h1}, \eq{h2} (or equivalently \eq{hc}),
together with the projection condition \eq{it}.


\subsection{Conditions From $\delta\chi=0$}


The analysis for this case is identical to that given in
\cite{Cariglia:2004kk}, so we will skip the details, referring to
this paper. Multiplying \eq{s2} with ${\bar\e}^B$ and
${\bar\e}^B\C_{\lambda\tau}$ gives four equations which can be
satisfied by
 \be V^\mu \partial_\mu \vf =0\ , \la{c1}\ee
and parametrizing $G^-$ as
\be e^{\ft12\vf}\,G^- = \ft12(1- \star) \left[V\wedge e^-\wedge d\vf
+V\wedge K\right]\ ,\la{c2} \ee
where $\star$ is the Hodge-dual,  $K=\ft12 K_{ij}\, e^i\wedge e^j$
is self-dual. In fact, these two conditions are the necessary and
sufficient conditions for satisfying $\delta\chi=0$.


\subsection{Conditions From $\delta\psi_\mu=0$}


Multiplying \eq{s1} with  ${\bar \epsilon}\C_\nu$, we find
\be \nabla_\mu V_\nu = -\ft12 e^{\ft12\vf}\,G^+_{\mu\nu\rho}V^\rho\
,  \la{killing} \ee
which implies that $V^\mu$ is a Killing vector. Similarly,
multiplying \eq{s1} with ${\bar \epsilon}\C_{\nu\rho\sigma}$ gives
an expression for $\nabla_\sigma  X^r_{\mu\nu\rho}$. Using
\eq{killing} one finds that this expression is equivalent to
\be D_\mu I^r_{ij} = e^{\ft12\vf} G^{+k}{}_{\mu [i}\,I^r_{j]k}\
,\la{di} \ee
where $D_\mu I^r \equiv \nabla_\mu I^r+\e^{rst} Q_\mu^s I^t$. One can use \eq{di}
to fix the composite $Sp(1)_R$ connection as follows
\be Q_\mu^r= \ft14 e^{\vf}G^{(+)}_{\mu ij} I^{rij} -\ft18 \e^{rst}
I^{sij} \nabla_\mu I^t_{ij}\ .\la{cc} \ee

Manipulations similar to those in \cite{Cariglia:2004kk} shows that,
using \eq{it} and \eq{killing}, the variation $\delta\psi_\mu=0$ is
directly satisfied, with $\epsilon$ constant, in a frame where
$I^r_{ij}$ are constants.

In summary, the necessary and sufficient conditions for
$\delta\psi_\mu=0$ are \eq{killing}, \eq{di}, together with the
projection condition \eq{it}.


\section{Integrability Conditions for the Existence of a Killing Spinor}


Assuming the Killing spinor conditions derived in the previous
section, the attendant integrability conditions can be used to show
that certain field equations are automatically satisfied. Since the
field equations are complicated second order equations, it is
therefore convenient to determine those which follow from the
integrability, and identify the remaining equations that need to be
satisfied over and above the Killing spinor conditions.

Let us begin by introducing the notation
\be \delta\psi_\mu={\widetilde D}_\mu\e\ ,\qquad \delta\chi=\ft14
\Delta\e\ ,\qquad \delta\lambda^I=e^{-\ft12\vf} \Delta^I\e\ ,\qquad
\delta\psi^a=\Delta^{aA}\e_A\ , \ee

for the supersymmetry variations and \be R_{\mu\nu}= J_{\mu\nu}\
,\quad \square\vf=J\ ,\quad D_\mu(e^{\ft12\vf}F^{I\mu\nu})=J^{I\nu}\
,\quad D_\mu P^{\mu aA} =J^{aA}\ ,\ee for bosonic field equations.
Then we find that \bea \C^\mu[{\widetilde D}_\mu, \Delta^I]\e^A &=&
2 \left[
D_\mu(e^{\ft12\vf}F^{I\mu\nu})-J^{I\nu}\right]\C_\nu\e^A\nn\w2
&& + e^{\ft12\vf} \left(D_\mu
F^I_{\nu\rho}\right)\C^{\mu\nu\rho}\e^A -8\C^\mu \left(D_\mu
C^{IAB}+2C^{Ia(A} P_{\mu a}{}^{B)}\right)\e_B \nn\w2
&&-2[\Delta,\Delta^I]\e^A + 2e^{\ft12\vf}
F^I_{\mu\nu}\C^{\mu\nu}\, (\delta\chi^A)
+16C^{IaA}\,(\delta\psi_a)\nn \ ,\w2
&&+8e^{\ft12\vf} f^{IJK}A_\mu^J \C^\mu (\delta\lambda^{KA}) \ ,
\la{e1}\w2
\C^\mu [{\widetilde D}_\mu,\Delta^{aA} ] \e_A &=& \left(D_\mu P^{\mu
aA} -J^{aA}\right)\e_A \nn\w2
&& +\C^{\mu\nu}\left(D_\mu P_\nu^{aA}-\ft12
F_{\mu\nu}^IC^{IaA}\right) \e_A \nn\w2
&&-4C^{IaA} (\delta\lambda^I_A) -\ft1{24}
e^{\ft12\vf}G_{\mu\nu\rho}\C^{\mu\nu\rho}\, (\delta\psi^a)\ ,
\la{e2}\w2
\C^\mu[{\widetilde D}_\mu,\Delta]\e_A &=& \left(\square\vf
-J\right)\e_A -\ft12 e^{-\ft12\vf} D_\mu (e^{\vf}
G^\mu{}_{\nu\rho})\, \C^{\nu\rho} \e_A\nn\w2
&& -\ft16 e^{\ft12\vf} \C^{\mu\nu\rho\sigma} \left(\nabla_\mu
G_{\nu\rho\sigma} -\ft34 F^I_{\mu\nu} F^I_{\rho\sigma}\right)
\e_A\nn\w2
&&
-\left(e^{\ft12\vf}F^I_{\mu\nu}\C^{\mu\nu}\e_{AB}+8C^I_{AB}\right)\,
\delta\lambda^{IB} +\ft16
e^{\ft12\vf}G_{\mu\nu\rho}\C^{\mu\nu\rho}\,(\delta\chi_A)\
,\la{e3}\w2
\C^\nu [{\widetilde D}_\mu,{\widetilde D}_\nu]\e^A &=& \ft12 \left(
R_{\mu\nu} -J_{\mu\nu}\right)\C^\nu \e^A  +\ft1{16}
e^{-\ft12\vf}\nabla^\nu (e^{\vf}
G_{\nu\rho\sigma})\,\C^{\rho\sigma}\C_\mu \e^A\nn\w2
&& +\ft1{48} e^{\ft12\vf}\C^{\rho\sigma\lambda\tau}\C_\mu
\left(\nabla_\rho G_{\sigma\lambda\tau} -\ft34
F^I_{\rho\sigma}F^I_{\lambda\tau} \right)\e^A\nn\w2
&& + \left( Q_{\mu\nu}^{AB}+ F_{\mu\nu}^I C^{IAB} -2 P_{[\mu}^{aA}
P_{\nu] a}{}^B\right)\C^\nu \e_B\nn\w2
&&+\ft12\left[\partial_\mu\vf +\ft1{12} e^{\ft12\vf}
G_{\nu\rho\sigma}\C^{\nu\rho\sigma}\C_\mu\right]\,\delta \chi^A +2
P_\mu^{aA} (\delta \psi_a )\nn\w2
&& -\ft18 e^{\ft12\vf}\left[ (\C^{\nu\rho}\C_\mu-4\delta_\mu^\nu
\C^\rho)F_{\nu\rho}^I \e^{AB}- \C_\mu C^{IAB}\right]
\,\delta\lambda^I_B\la{e4} \ .\eea

If one makes the ansatz for the potentials directly , then the
Bianchi identities and the relations \eq{dc} and \eq{id1}--\eq{id3}
are automatically satisfied. Otherwise, all of these equations must
be checked. Assuming that these are satisfied, from \eq{e1} it
follows that the Yang-Mills field equation $K_\mu=0$, {\it  except
for $K_+=0$}, is automatically satisfied, as can be seen by
multiplying $K_\mu\C^\mu\e^A=0$ by ${\bar\e}^B$ and $K_\nu\C^\nu $,
recalling $\C^+\e=0$ and further simple manipulations. Similarly,
from \eq{e2} it follows that the hyperscalar field equation
$K^{aA}=0$ is automatically satisfied as can be seen by multiplying
$K^{aA}\e_A=0$ by ${\bar\e}_B\C^\mu$. Finally, from \eq{e3} and
\eq{e4}, it follows that the dilaton and Einstein equation
$E_{\mu\nu}=0$, {\it except $E_{++}=0$}, are automatically
satisfied, provided that we also impose the $G$-field equation. This
can be seen by multiplying $E_{\mu\nu}\C^\nu\e_A=0$ with
${\bar\e}_B$ and $E_{\mu\rho}\C^\rho$ and simply manipulations that
make use of $\C^+\e=0$.

In summary, once the Killing spinor conditions are obeyed, all the
field equations are automatically satisfied as well, except the
following,
\be R_{++}=J_{++}\ ,\qquad D_\mu(e^{\ft12\vf} F^{I\mu}{}_{+})=J^I_+\
,\quad\quad D_\mu(e^{\vf}G^{\mu\nu\rho})=0\ , \label{remaining} \ee
and the Bianchi identities $DF^I=0$ and $dG=\ft12 F^I\wedge F^I$.

It is useful to note that in the case of gravity coupled to a
non-linear sigma model, the scalar field equation follows from the
Einstein's equation and the contracted Bianchi identity only when
the scalar map is a submersion (i.e. when the rank of the matrix
$\del_\m\phi^\a$ is equal to the dimension of the scalar manifold).
In our model, however, the scalar field equation is automatically
satisfied as a consequence of the Killing spinor integrability
conditions, without having to impose such requirements. This is all
the more remarkable given the fact that there are contributions to
the energy-momentum tensor from fields other than the scalars.

Finally, in analyzing the set of equations summarized above for
finding a supersymmetric solution, it is convenient to parametrize
the metric, which admits a null Killing vector, in general as
\cite{Gutowski:2003rg}
\be ds^2= 2H^{-1}(du+\beta)\left(dv+\omega+{{\cal F}\over
2}(du+\beta)\right) +H ds_B^2\ , \label{gm}\ee
with
\bea e^+ &=&H^{-1}(du+\b)\ ,\nn\w2
e^- &=& dv+\omega+\ft12 {\cal F} H e^+\ ,\nn\w2
e^i &=& H^{1/2} {\tilde e}_\a{}^i dy^\a\ , \eea
where $ds_B^2 =h_{\a\b}dy^\a dy^\b$ is the metric on the base space
${\cal B}$, and we have $\beta=\beta_\a dy^\a$ and $\omega=\omega_\a
dy^\a$ as $1$-forms on ${\cal B}$.  These quantities as well as the
functions $H$ and ${\cal F}$ depend on $u$ and $y$ but not on $v$.
Now, as in \cite{Gutowski:2003rg}, defining the $2$-forms on ${\cal
B}$ by
\be {\tilde J}^r = H^{-1}I^r\ ,\ee
these obey
\be ({\tilde J}^r)^\a{}_\g\,({\tilde J}^s)^\gamma{}_\b = \e^{rst}
({\tilde J}^t)^\a{}_\b - \delta^{rs}\delta^\a_\b\ ,\label{hks2} \ee
where raising and lowering of the indices is understood to be made
with $h_{\a\b}$. Note that the index $\a=1,...,4$ labels the
coordinates $y^\a$ on the base space ${\cal B}$. This should not be
confused with the index $\a=1,...,n_H$ that labels the coordinates
$\phi^\a$ of the scalar manifold!

A geometrically significant equation satisfied by ${\tilde J}^r$ can
be obtained from \eq{di}, and with the help of \eq{killing} it takes
the form \cite{Cariglia:2004kk},
\be {\tilde\nabla}_i {\tilde J}^r_{jk} +\epsilon^{rst} Q_i^s {\tilde
J}^t_{jk}- \beta_i {\dot {\tilde J}}^r_{jk}-{\dot\beta}_{[j}{\tilde
J}^r_{k]i} +\d_{i[j} {\dot \beta}^m {\tilde J}^r_{k]m}=0\
,\label{djb} \ee
where ${\tilde\nabla}_i$ is the covariant derivative on the base
space ${\cal B}$ with the metric $ds_B^2$ and $\dot\b \equiv
\partial_u\b$.


\section{The Dyonic String Solution}


For the string solution we shall activate only four hyperscalars,
setting all the rest equal to zero. In the quaternionic notation of
Appendix B, this means
\be t= \left(
            \begin{array}{c}
              \phi \\
              0 \\
              \vdots \\
             0 \\
            \end{array}
          \right) \la{fvec}
\ee
In what follows we shall use the map
\be \phi=\phi^{A'A}= \phi^\a (\sigma_{\a})^{A'A}\ , \la{qdef}\ee
where $ \sigma_{\a}=(1,-i \vec{\sigma} )$ are the constant van der
Wardeen symbols for $SO(4)$. Moreover, we shall chose the gauge
group $K$ such that
\be T^{I'}\, t =0\ .\la{tprime} \ee
This condition can be easily satisfied by taking  $K$ to be a
subgroup of $Sp(n_H-1)$ which evidently leaves $t$ given in
\eq{fvec} invariant. Finally, we set
\be A_\mu^{I'}=0\ . \la{aprime}\ee
Then, supersymmetry condition \eq{fi} in $I'$ direction is satisfied
by setting ${\widetilde F}^{I'}=0=\omega^{I'}$ and noting that
$C^{I'r}=0$ in view of \eq{tprime} (see \eq{ci}). The supersymmetry
condition \eq{h2} is also satisfied along the directions in which
the hyperscalars are set to zero. Therefore, the model effectively
reduces to one in which the hyperscalars are described by
$Sp(1,1)/Sp(1)\times Sp(1)$, which is equivalent to a 4-hyperboloid
$H_4=SO(4,1)/SO(4)$.

Using \eq{qdef} in the definition of $D_\mu t$
given in \eq{ddf}, we obtain
\be D_\m\phi^{\a}= \partial_\mu \phi^{\a}-\ft12 A_\mu^r
(\rho^r)^{\a}{}_{\b}\,\,\phi^{\b} \ , \ee
where the 't Hooft symbols $\rho^r$ are constant matrices defined as
\be \rho^r_{\a\b}= {\rm tr}\,(\sigma_{\a}\, T^r\,{\bar\sigma}_{\b})
\ . \ee
These are anti-self dual and their further properties are given in
Appendix A.

For the metric we choose
\be \beta=0\ ,\qquad \omega=0\ ,\qquad {\cal F}=0\ ,  \qquad
h_{\a\b}=\Omega^2 \delta_{\a\b}\ , \ee
in the general expression \eq{gm}, so that our ansatz takes the form
\be ds^2=2 H^{-1} \,du dv + H ds_B^2\ ,\quad\quad ds_B^2=\Omega^2
dy^\a dy^\b\delta_{\a\b}\ , \label{ds4}\ee
where $\Omega$ is a function of $y^2\equiv y^\a y^\b\d_{\a\b}$. We
also choose the null basis as
\be e^+=V= H^{-1} du\ ,\quad\quad e^-=dv\ . \ee
Thus, $V^\mu\partial_\mu=\partial /\partial v$. Moreover, in the
rest of this section, {\it we shall take all the fields to be
independent of $u$ and $v$}. Given that $\beta=0$, it also follows
from \eq{djb} that
\be {\tilde\nabla}_i {\tilde J}^r_{jk} +\epsilon^{rst} Q_i^s {\tilde
J}^t_{jk}=0\ . \label{djv} \ee
Next, in the general form of $G^{(-)}$ given in \eq{c2}, we choose
\be K=0 \ . \ee
Then, from \eq{c2} and \eq{killing} we can compute all the
components of $G^+$ and $G^-$, which yield for $G=G^++G^-$ the
result
\be G=e^{-\vf/2} \left(e^+\wedge e^-\wedge d\vf_+ + \star_4\,
d\vf_-\right)\ , \ee
where $\star_4$ refers to Hodge dual on the transverse space with
metric
\be ds_4^2= H ds_B^2\ ,\label{4dm} \ee
and we have defined
\be
 \vf_\pm  \ := \ \pm \ft12 \vf +\ln\,H\ . \ee
Next, we turn to the supersymmetry condition \eq{hc} in the
hyperscalar sector. With our ansatz described so far, it can now be
written as
\be D_i\phi^{\ua} = ({\tilde J}^r)_i{}^j\,(J^r)_{\ub}{}^{\ua}\,
D_j\phi^{\ub}\ , \la{hc4} \ee
where
\be D_i\phi^{\ua} \equiv  D_i\phi^\a\,V_\a{}^{\ua}\ , \ee
and $ V_\a{}^{\ua}$ is the vielbein on $H_4$, and the above
equations are in the basis
\be {\tilde e}^i = \d_\a^i\, \Omega\, dy^\a\ ,\label{bb} \ee
referring to the base space ${\cal B}$. We also note that
\be J^r_{\ua\ub}= \rho^r_{\a\b}\,\d^\a_{\ua} \,\d_{\ub}^\b\ , \ee
which follows from rom \eq{jr} and \eq{av}. Recall that the 't Hooft
matrices $\rho^r_{\a\b}$ are constants. Next, we choose the
components of ${\tilde J}^r_{ij}$ to be constants and  make the
identification
\be {\tilde J}^r=J^r\ .\label{jtilde}\ee
Using the quaternion algebra, we can now rewrite \eq{hc4} as
\be D_i\phi_{\ub}=
\left(\d_{i\ua}\d_{j\ub}-\d_{j\ua}\d_{i\ub}-\e_{ij\ua\ub}\right)\,D_j\phi_{\ua}\
. \la{dfi}\ee
Symmetric and antisymmetric parts in  $i$ and $\ub$ give
\bea
&& D_i \phi^i = 0\ ,\ \ \ \ \ \phi^i\equiv
\phi^{\ua}\,\d_{\ua}^i\ ,\label{ss1}\w2
&& D_i\phi_j -D_j\phi_i = -\e_{ijk\ell} D_k \phi_\ell\ .
\label{ss2}\eea
To solve these equations, we make the ansatz
\be \phi^\a=f y^\a\, ,\quad\quad A^r_\a=g\, \rho^r_{\a\b}\,y^\b\ ,
 \label{af}\ee
where $f$ and $g$ are functions of $y^2$. This ansatz, in
particular, implies that the function $\omega^r$ arising in the
general form of $F^r$ given in \eq{fi} vanishes. Assuming that the
map $\phi^\a$ is 1-1, one can actually use diffeomorphism invariance
to set (at least locally) $f=1$. However, since we have already
fixed the form of the metric as in \eq{ds4}, chosen a basis as in
\eq{bb}, and identified the components of the quaternionic
structures ${\tilde J}^{r}_{ij}$ referring to this orthonormal
basis, the reparametrization invariance has been lost. Therefore it
is important to keep the freedom of having an arbitrary function in
the map \eq{af}.

Using \eq{af} we find that \eq{ss2} is identically satisfied and \eq{ss1} implies
\be g=\fr{4 f' y^2+8f}{3fy^2} \ ,\label{son1}\ee
where prime denotes derivative with respect to argument, i.e. $y^2$.
Next, the computation of the Yang-Mills field strength from the
potential \eq{af} gives the result
\bea
&&F^r = F^{r(+)}+F^{r(-)}\ ,\qquad F^{r\pm}= \pm \star_4 F^{r\pm}\ ,
\w2
&& F^{r(-)}_{\a\b} = (-2g-g'y^2+\ft12g^2y^2)\,\rho^r_{\a\b}\ ,\nn\w2
&& F^{r(+)}_{\a\b} \equiv {\widetilde F}^r_{\a\b}= (2g'+
g^2)\,\left( 2 y_{[\a}y^\d\,\rho^r_{\b]\d} +\ft12
y^2\,\rho^r_{\a\b}\right)\ .\nn \eea
Comparing these results with the general form of $F^I$ given in
\eq{fi}, we obtain
\be e^{\vf_-}= {\eta\over \Omega^2}\ , \label{son2}\ee
where
\be \eta \equiv \left(g'y^2+2g-\ft12 g^2y^2\right)(1-f^2y^2)\ .
\label{eta}\ee
Here we have used the fact that $C^{r,s}=\d^{rs}/(1-\phi^2)$ as it
follows from the formula \eq{crr}. Finally using the composite
connection \eq{qsp1} in \eq{djv} we obtain
\be \fr{\O '}{\O}= \fr{(2f^2-g)}{2(1-f^2y^2)}\ . \label{son3}\ee
This equation can be integrated with the help of \eq{son1}, yielding
\be \Omega = {b\over y^2}\left( {1-f^2 y^2\over f^2
y^2}\right)^{1/3}\ ,\label{omega} \ee
where $b$ is an integration constant. One can now see that all
necessary and sufficient conditions for the existence of a Killing
spinor on this background are indeed satisfied. As shown in the
previous section, the integrability conditions for the existence of
a Killing spinor imply all field equations except \eq{remaining} and
the Bianchi identities on $F^I$ and $G$. It is easy to check that
\eq{remaining} is identically satisfied by our ansatz, except for
the $G$-field equation. Furthermore, the Yang-Mills Bianchi identity
is trivial since we give the potential. Thus, the only remaining
equations to be checked are the $G$-Bianchi identity and the
$G$-field equation. To this end, it is useful to record the result
\be {\epsilon^{\a\b\gamma\d}\over \sqrt {g_4}} F_{\a\b}^r
F^r_{\gamma\delta}= {16 Q'\over y^2 H^2\Omega^4}\ , \ee
where $g_4$ is the determinant of the metric for the line element
$ds_4^2$, and
\be Q \equiv  (gy^2)^2 (gy^2-3) +c\ , \ee
where $c$ is an integration constant. Interestingly, this term is
proportional to the sum of of $F^2$ and $C^2$ terms that arise in
the dilaton field equation, up to an overall constant.

We now impose the $G$-field equation $d (e^{\vf} \star G)=0$ and the
$G$-Bianchi identity $dG=\ft12 F^r\wedge F^r$. The $G$-field
equation gives\
\be \square_4 \vf_+ +\ft12 \partial_\a\, \vf \partial^\a \vf_+ \ =\
0\ , \label{son4} \ee
and the $G$-Bianchi identity amounts to
\be \square_4 \vf_- -\ft12 \partial_\a \vf\, \partial^\a \vf_-  \ =\
{-2Q'\over y^2 H^2\Omega^4}\ ,\label{son5} \ee
where the Laplacian is defined with respect to the metric \eq{4dm}.
These equations can be integrated once to give
\be \vf_+'= {\nu e^{-\vf}\over (y^2)^2 \eta}\ ,\quad\quad \vf_-'=
{(\lambda-\frac12 Q)\over (y^2)^2 \eta}\ , \label{son6}\ee
where $\nu,\lambda$ are the integration constants, $c$ has been
absorbed into the definition of $\lambda$, and \eq{son2} has been
used in the form $H\Omega^2=\eta e^{\vf/2}$. These equation can be
rewritten as
\bea
\left(e^{\vf_+}\right)' &=& {\nu\over b^2}\left(  {f^2 y^2\over
1-f^2 y^2}\right)^{2/3}\ ,\label{fplus}\w2
\left(e^{\vf_-}\right)' &=& {\lambda-\frac12 Q \over b^2}\left( {f^2
y^2\over 1-f^2 y^2}\right)^{2/3}\ ,\label{fminus} \eea
by recalling $\vf=\vf_+ - \vf_-$, exploiting \eq{son2} and using the
solution \eq{omega} for $\Omega$. It is important to observe that
the second equation in \eq{son6}, has to be consistent with
\eq{son2}. Differentiating the latter and comparing the two
expressions, we obtain a third order differential equation for the
function $f$:
\be\eta'- \left({2f^2 -g\over 1-f^2y^2}\right)\eta= {\lambda-\frac12
Q\over (y^2)^2}\ . \label{fin2}\ee
In summary, any solution of this equation for $f$ determines also
the functions $(\vf,H,\Omega, g)$, and therefore fixes the solution
completely. This is a highly complicated equation, however, and we
do not know its general solution at this time. Nonetheless, it is
remarkable that an ansatz of the form
\be f= {a\over y^2}\ , \ee
with $a$ a constant, which gives $g=4/(3y^2)$ from \eq{son1}, does
solve \eq{fin2}, and moreover, it fixes the integration constant
\be \lambda=-\ft43 \ . \ee
Furthermore, it follows from \eq{omega}, \eq{son2}, \eq{eta} and
\eq{fplus} that
\be \Omega = {b\over y^2}\, h^{1/3}\ ,\qquad e^{\vf_-} =
\left({2a\over 3b}\right)^2 h^{1/3}\ ,\qquad e^{\vf_+}= 3\nu
\left({a\over b}\right)^2 h^{1/3} +\nu_0\ , \label{defs}\ee
where $\nu_0$ is an integration constant and
\be h \equiv {y^2\over a^2}-1\ . \ee
Thus, the full solution takes the form
\bea ds^2 &=& e^{-\ft12\vf_+}e^{-\ft12\vf_-}(-dt^2+dx^2) +
e^{\ft12\vf_+}e^{\ft12\vf_-} \left({b\over
y^2}\right)^2\,h^{2/3}\,dy^\a dy^\b\,\d_{\a\b}\ ,\label{smet}\w2
e^\vf&=& e^{\vf_+}/e^{\vf_-}\ ,\quad\quad \phi^\a= {a y^\a\over
y^2}\ ,\w2
A_\a^r &=& {4\over 3 y^2}\,\rho^r_{\a\b} y^\b\ ,\w2
G_{\a\b\gamma} &=& {8\over 27 (y^2)^2}\,\e_{\a\b\gamma\d}\,y^\d\ ,
\qquad G_{+-\a} = -\partial_\a e^{-\vf_+}\ ,  \eea
with $\vf_\pm$ given in \eq{defs}. The form of $h$ dictates that
$a^2 < y^2 <\infty$, covering outside of a disk of radius $a$. The
hyperscalars map this region into  $H^4$ which can be viewed as the
interior of the disk defined by $\phi^2 < 1$. These scalars are
gravitating in the sense that their contribution to the energy
momentum tensor, which takes the form $(\tr P_iP_j-\ft12 g_{ij}\tr
P^2)$, does not vanish since the solution gives
\be P_i^{A'A}= {a\over 3 y^2\left(1-\frac{a^2}{y^2}\right)}
\left(\d_i^\a - 4 {y_iy^\a\over y^2}\right)\, \s_\a^{A'A}\ . \ee

It is possible to apply a coordinate transformation and map the base
space into the disc by defining
\be z^\a\equiv \fr{a y^\a}{y^2}. \ee
In $z^\a$ coordinates the solution becomes
\bea
ds^2 &=& e^{-\ft12\vf_+}e^{-\ft12\vf_-}(-dt^2+dx^2) + L^2
e^{\ft12\vf_+}e^{\ft12\vf_-} \, h^{2/3}\, (dr^2 +r^2 d\Omega_3^2)\
\label{smetz}\w2
e^\vf&=& e^{\vf_+}/e^{\vf_-}\ ,  \label{zfi}\w2
G &=& \ft{8}{27}\, \Omega_3 - dt\wedge dx \wedge de^{-\vf_+}\ ,\w2
A^r &=& \ft23\, r^2 \s^r_R \ , \w2
\phi^\a &=& z^\a \ , \label{zha} \eea
where
\bea && r= \sqrt{z^\a z^\b \d_{\a\b}}\ ,\qquad  \Omega_3 = \s^1_R
\wedge \s^2_R \wedge \s^3_R\ , \qquad h={1\over r^2}-1\ , \w2
&& e^{\vf_+}= {3\nu h^{1/3}\over  L^2}  +\nu_0\ ,\qquad e^{\vf_-} =
{4 h^{1/3}\over 9L^2}\ , \label{har}\eea
and $L\equiv b/a$. Here, $\s^r_R$ are the right-invariant one-forms
satisfying
\be d\s^r_R = \ft12 \e^{rst}\, \s^s_R\wedge \s^t_R\ ,\ee
and $\Omega_3$ is the volume form on $S^3$. We have also used the
definitions
\be    z^\a= r\, n^\a\ ,\qquad n^\a n^\b\delta_{\a\b}=1\ , \ee
where $dn^\a$ are orthogonal to the unit vectors $n^\a$ on the
$3$-sphere, and satisfy
\be dn^\a=\ft12 \rho^{r\a}{}_\b\,\s^r_R\,n^\b\ ,\qquad dn^\a dn^\b
\d_{\a\b} = \ft14 d\Omega_3^2\ . \ee
Given the form of $A^r$, it is easy to see that the Yang-Mills
$2$-form $F^r=dA^r-\ft12 \e^{rst} A^s\wedge A^t$  is not
(anti)self-dual, as it is given by
\be   F^r=\ft43\,rdr\wedge \s^r_R +\ft13 r^2 \left(1-\ft23
r^2\right)\,\e^{rst}\s^s_R\wedge \s^t_R\ . \ee
The field strength  $P_i^{A'A}$ on the other hand, takes the form
\be P_i^{A'A}= {1\over 1-r^2}\,\left[ (1-\ft23 r^2) \d_i^\a +\ft23
r^2 n_i n^\a\right]\, \s_\a^{A'A}\ . \ee
We emphasize that, had we started with the identity map
$\phi^\a=z^\a$ from the beginning, the orthonormal basis in which
${\tilde J}^r_{ij}$ are constants would be more complicated than the
one given in \eq{bb}. Consequently, \eq{son3} would change since it
uses \eq{djv} that requires the computation of the  spin connection
in the new orthonormal basis.

\section{Properties of the Solution}

\subsection{Dyonic Charges and Limits}

To begin with, we observe that the solution we have presented above
is a dyonic string with with {\it fixed} magnetic charge given by
\be Q_m=\int_{S^3} G = \frac{8}{27}\, vol_{S^3}\ . \ee
The electric charge, however, turns out to be proportional to the
constant parameter $\nu$ as follows:
\be Q_e=\int_{S^3} \star e^\vf G = 2\nu\, vol_{S^3}\ . \ee
Next, let us compare our solution with that of \cite{Guven:2003uw}
where a dyonic string solution of the $U(1)_R$ gauged model in the
absence of hypermatter has been obtained. We shall refer to this
solution as the GLPS  dyonic string. To begin with, the GLPS
solution has two harmonic functions with two arbitrary integration
constants, as opposed to our single harmonic function $h$ with a
fixed and negative integration constant. In our solution, this is
essentially due to the fact that we have employed an identity map
between a hyperbolic negative constant curvature scalar manifold and
space transverse to the string worldsheet.

Next, the transverse space metric $ds_4^2$ in the GLPS solution is a
warped product of a {\it squashed} $3$-sphere with a real line,
while in our solution it is conformal to $R^4$. In GLPS solution the
deviation from the round $3$-sphere is proportional to a product of
$U(1)_R$ gauge constant and monopole flux due to the $U(1)_R$ gauge
field. Thus, assuming that we are dealing with a gauged theory, the
round $3$-sphere limit would require the vanishing of the monopole
flux, which is not an allowed value in GLPS solution.

As for the $3$-form charges, the electric charge is arbitrary in the
GLPS as well as our solution. However, while the magnetic charge in
the GLPS solution is proportional to $k\xi/g_R$ where $k$ is the
monopole flux, $g_R$ is the $U(1)_R$ coupling constant and $\xi$ is
the squashing parameter, and therefore arbitrary, in our solution
the magnetic charge is fixed in Planckian units and therefore it is
necessarily non-vanishing. This is an interesting property of our
solution that results from the interplay between the sigma model
manifold whose radius is fixed in units of Plank length, which is
typical in supergravities with a sigma model sector, and the four
dimensional space transverse to the the string worldsheet.

Our solution has $SO(1,1)\times SO(4)$ symmetry corresponding to
Poincar\'e invariance in the string world-sheet and rotational
invariance in the transverse space\footnote{It is clear that if one
makes an $SO(4)$ rotation in $z^\a$ coordinates, the same
transformation should be applied to hyperscalars and 't Hooft
symbols $\r^r_{\a\b}$ to preserve the structure of the solution.}.
The metric components exhibit singularities at $r=0$ and $r=1$. Too
see the coordinate invariant significance of these points, we
compute the Ricci scalar as
\be R= { 48 (\Delta+\mu_0)^2+\mu_0^2 \over r^6 \left({\Delta\over
3\nu}\right)^{\ft{17}{18}}(\Delta+\mu_0)^{\ft52}}\ , \ee
where $\Delta\equiv 3\nu ({1\over r^2}-1)$ and $\mu_0\equiv \nu_0
L^2$.
We see that, near the boundary $r\to 1$, the Ricci scalar diverges,
and there is a genuine singularity there. Near the origin $r=0$,
however, the situation depends on the parameter $\nu$. If $\nu\ne
0$, then as $r\to 0$ the Ricci scalar approaches the constant value
$8/\sqrt{3\nu}$. The metric is perfectly regular in this limit, and
indeed, we find that it takes the form
\be ds^2 \to  {L^2\over R_0^2}\, r^{2/3} (-dt^2+dx^2) + {R_0^2
dr^2\over r^2} + R_0^2 d\Omega_3^2\ , \ee
which is $AdS_3 \times S_3$ with $R_0= \sqrt{4\nu/3}$. This is to be
contrasted with the GLPS solution which approaches the product of
$AdS_3$ with a squashed $3$-sphere.

The $r=0$ point can be viewed as the horizon, and as is usually the
case, our solution also has a factor of two enhancement of
supersymmetry near the horizon. This is due to the fact that the
condition \eq{susy1}, which reads $H^{-1}\Gamma^+\epsilon=0$ has to
be relaxed since $H^{-1}$ vanishes in in the $r\to 0$ limit. Note,
however, that our solution at generic point has $1/8$ supersymmetry
to begin with, as opposed to $1/4$ supersymmetry of the GLPS
solution.

For $\nu=0$, the $r\to 0$ limit of the metric is
\be ds^2 \to {3L\over 2\sqrt {\nu_0}}\, r^{1/3}(-dt^2+dx^2) +
{2L\sqrt{\nu_0}\over 3}\, r^{-5/3} ({dr^2}+r^2d\Omega_3^2) \ , \ee
Defining furthermore $du=dr/r^{5/6}$ the metric becomes
\be ds^2\sim
u^{2}(-dt^2+dx^2+d\O_3^2)+ du^2. \ee
Ignoring $x$ and $\O_3$ directions, this describes the Rindler wedge
which is the near horizon geometry of the Schwarzcshild black hole.
The ``horizon'', which has the topology $R\times \O_3$, shrinks to
the zero size at $u=0$ and this gives the singularity in the dyonic
string.

Next, consider the boundary limit in which $ r\to 1$. First,
assuming that $\nu_0 \ne 0$, we find that in the limit $r\to 1$ the
metric takes the form
\be ds^2 \sim {1\over u^{1/3}} \left( -dt^2 + dx^2 + u^4 ( \,du^2
 +{1\over u^2}\,d\Omega_3^2) \right) \qquad \mbox{for} \ \ \nu_0\ne 0\ ,\ee
where we have defined the coordinate $u=h^{1/2}$ and rescaled the
string worldsheet coordinates by a constant.  For $\nu_0=0$, on the
other hand, the $r\to 1$ limit of the metric is given by
\be ds^2 \sim {1\over u^{2/3}} \left( -dt^2 + dx^2 \right) + u^4
\left( \,du^2
 +{1\over u^2}\,d\Omega_3^2 \right) \qquad \mbox{for} \ \ \nu_0 = 0\ ,\ee
where, again, we have defined $u=h^{1/2}$ and rescaled coordinates
by constants.


\subsection{Coupling of Sources}


Since the solution involves the harmonic function $h$, there is also
a possibility of a delta function type singularity at the origin
since
\be \del_\a\del^\a\, h=-4\pi^2 \d(\vec{z})\, .\label{ds} \ee
The presence of such a singularity requires addition of extra
sources to supergravity fields to get a proper solution. As it is
not known how to write down the coupling of a dyonic string to
sources, and as we cannot turn off the magnetic charge, we consider
the coupling of the magnetic string to sources. Thus setting
$\nu=0$, from \eq{smetz}, \eq{zfi} and \eq{zha} the dangerous fields
that can possibly yield a delta function via \eq{ds} are the metric,
the dilaton $\phi$ and the three form field $G$. Indeed from
\eq{zha} we see that \be dG\sim \d(\vec{z})\, dz^1\wedge dz^2\wedge
dz^3\wedge dz^4\label{sin1}\, , \ee therefore extra (magnetically
charged) sources are needed for $G$ at $\vec{z}=0$. For the dilaton
we find that the candidate singular term near $\vec{z}=0$ behaves as
\be \square \varphi\sim \,z^{11/3}\,\d(\vec{z})\to 0\ ,
\label{sinf}\ee
thus  there is
no problem  at $\vec{z}=0$. Finally for the Ricci tensor expressed
in the coordinate basis we find
\bea R_{tt}&=&-R_{xx}\sim z^{4} \d(\vec{z}) \to 0\ ,  \label{sin}\w2
R_{\a\b}&\sim& \,z^2 \d(\vec{z})\,\d_{\a\b} \to 0\ . \label{sin2} \eea
Contracting with the metric one can see that the possible singular
part in the Ricci scalar becomes
\be R\sim \,z^{11/3}\,\d(\vec{z})\to 0\ , \ee
and thus there appears no extra delta function singularity.

The above results can be understood by coupling to supergravity
fields a magnetically charged string located at $r=0$ with its
action given by
\be S = -\int d^2\s e^{\varphi/2} \sqrt {-\gamma} +\int {\widetilde
B}\ , \ee
where $\gamma$ is the determinant of the induced worldsheet metric
and ${\widetilde B}$ is the 2-form potential whose field strength is
dual to $G$. This coupling indeed produces exactly the behavior
\eq{sin1} in the Bianchi identity. The source terms in \eq{sinf} and
\eq{sin} are also produced, while the contribution to the right hand
side of \eq{sin2} vanishes identically (which does not causes a
problem since $z^2\delta({\vec z})$ vanishes at $z=0$ as well).

\subsection{Base Space as a Tear Drop}

In \eq{smetz} the four dimensional base space for our solution
\eq{smetz} is
\bea ds_B^2 &=& L^2 \left({1\over r^2}-1\right)^{2/3}\,
\left(dr^2+r^2d\O_3^2\right)\label{baseconf}\nn\w2
&=& {(1-r^2)^{8/3}\over 2r^{4/3}}\,ds_{H_4}^2\ , \eea
where $ds_{H_4}^2= 2 (dr^2 +r^2 d\Omega_3^2)/(1-r^2)^2$ is the
metric on $H_4$. Although the overall conformal factor blows at
$r=0$, the total volume of this space turns out to have a finite
value $(4\pi^3 L^4)/(9\sqrt{3})$. To that extent, our solution can
be viewed as the analog of the Gell-Mann-Zwiebach teardrop solution,
though the latter is regular at $r=0$ as well. The analogy with
Gell-Mann-Zwiebach tear-drop is also evident in the fact that the
scalar metric has been conformally rescaled by a factor that
vanishes at the boundary.

The curvature scalar of the base manifold is also singular at $r=0$,
as it is given by
\be R_{\cal B} = {16\over 3L^2}\,{1\over r^2}\,{r^{4/3}\over
(1-r^2)^{8/3}}\ . \ee
Since the total volume in the base space is finite, one would expect
that the singularity at $r=0$ can be reached by physical particles
at a finite proper time. We have checked that this is indeed the
case.

Another tear-drop like feature here is that the base space metric is
conformally related to that of $H_4$ which has negative constant
curvature, and that the curvature scalar of the bases space becomes
positive due to the conformal factor. This switching of  the sign is
crucial for satisfying Einstein equation in the internal direction,
just as in the case of 2-dimensional Gell-Zwiebach teardrop.

The base space ${\cal B}$ that emerges in the  $2+4$ split of the
$6D$ spacetime is quaternionic manifold, as it admits a quaternionic
structure. To decide whther it is Quaternionic Kahler (QK), however,
the standard definition  that relies on the holonomy group being
contained in $Sp(n)\times Sp(1) \sim SO(4)$ becomes vacuous in $4D$
since all $4D$ Riemann manifolds have holonomy group $sp(1)\times
Sp(1)$. Nonetheless, there exists a generally accepted and natural
definition of QK manifolds in four dimensions, which states that an
oriented $4D$ Riemann manifold is QK if the metric is self-dual and
Einstein (see \cite{Galicki} for a review). According to this
definition, our base space ${\cal B}$ is not QK since it is neither
self-dual nor Einstein.

\subsection{Reduction of Metric to Five Dimensions}

Finally, we would like to note the 5-dimensional metric that can be
obtained by a Kaluza-Klein reduction along the string direction. The
6-dimensional metric is parametrized in terms of the 5-dimensional
metric as \be ds_6^2=e^{2 \alpha \hat{\phi}} ds_5^2+ e^{2 \beta
\hat{\phi}} dx^2 \ee where $\beta=-3\alpha$ and $\hat{\phi}$ is the
Kaluza-Klein scalar. From \eq{smetz} one finds
\be ds_5^2=-e^{-\ft23 \vf_+}e^{-\ft23 \vf_-}\,dt^2+L^2
e^{\ft13\vf_+} e^{\ft13 \vf_-}h^{2/3}( dr^2+d\O_3^2), \label{bh} \ee
where the functions are still given in \eq{har}.

The metric \eq{bh} is singular at $r=0$. For $\nu=0$ looking at the
metric near the singularity one finds
\be ds_5^2\sim u^2(-dt^2+d\O_3^2)+du^2, \ee
where $du=dr/r^{7/9}$. The geometry is like the Rindler space but
the candidate spherical ``horizon'' shrinks to zero size at $u=0$
which produces a singularity. When $\nu\not=0$, one finds near $r=0$
that \be ds_5^2\sim -r^{8/9}dt^2+r^{-16/9}dr^2+r^{2/9}d\O_3^2 \ee
which is again singular at $r=0$. This singularity is resolved by
dimensional {\it oxidation} which is a well known feature of some
black-brane solutions \cite{Gibbons:1994vm}.


\section{Conclusions}


In this paper, we have derived the necessary and sufficient
conditions for the existence of a Killing spinor in $N=(1,0),\,6D$
gauge supergravity  coupled to a single tensor multiplet, vector
multiplets and hypermultiplets. This generalizes the analysis of
\cite{Gutowski:2003rg} and \cite{Cariglia:2004kk} by the inclusion
of the hypermatter. In our case as well, the existence of the
Killing spinor implies that the metric admits a null Killing vector.
This is in contrast to some other dimensions such as $D=4,5$ where
time-like and space-like Killing vectors arise in addition to the
null one. The Killing spinor existence conditions and their
integrability are shown to imply most of the equations of motion.
This simplifies greatly the search for exact solutions. The
remaining equations to be solved are (i) the Yang-Mills equation in
the null direction, (ii) the field equation for the $2$-form
potential, (iii) the Bianchi identities for the Yang-Mills curvature
and the field strength of the $2$-form potential, and (iv) the
Einstein equation in the double null direction. We parametrize the
most general form of a supersymmetric solution which involves a
number of undetermined functions. However, we do not write
explicitly the equations that these functions must satisfy. These
can be straightforwardly derived from the equations just listed.

The existence of a null Killing vector suggests a $2+4$ split of
spacetime, and search for a string solution, possibly dyonic. Such
solutions are already known but none of them involve any active
hyperscalar. As a natural application of the general framework
presented here, we have then focused on finding a dyonic string
solution in which the hyperscalars have been activated.

Indeed, we have found a $1/8$ supersymmetric such a dyonic string.
The activated scalars parametrize a $4$ dimensional submanifold of a
quaternionic hyperbolic ball of unit radius, characterized by the
coset $Sp(n_H,4)/Sp(n_H)\times Sp(1)_R$. A key step in the
construction of the solution is an identity map between the
$4$-dimensional scalar submanifold and internal space transverse to
the string worldsheet. The spacetime metric turns out to be a warped
product of the string worldsheet and a $4$-dimensional analog of the
Gell-Mann-Zwiebach tear-drop which is noncompact with finite volume.
Unlike the Gell-Mann-Zwiebach tear-drop, ours is singular at the
origin. There is also a delta function type singularity that comes
from the Laplacian acting on a harmonic function present in the
solution. This does not present any problem, however, as we place a
suitable source which produces contributions to the field equations
that balance the delta function terms.

An interesting property of our dyonic string solution is that while
its electric charge is arbitrary, its magnetic magnetic charge is
fixed in Planckian units, and hence it is necessarily non-vanishing.
This interesting feature results from the interplay between the
sigma model manifold whose radius is fixed in units of Plank length,
as it is the case in almost all supergravities that contain sigma
models, and the four dimensional space transverse to the the string
worldsheet through the identity map.

The tear-drop is quaternionic but not quaternionic Kahler, since its
metric is neither self-dual nor Einstein. The metric is conformally
related to that of $H_4$ which has negative constant curvature, and
its curvature scalar becomes positive due to the conformal factor.
This switching of  the sign is crucial for satisfying Einstein
equation in the internal direction, just as in the case of
2-dimensional Gell-Zwiebach teardrop.

We have also shown to have $1/4$ supersymmetric $AdS_3\times S^3$
near horizon limit where the radii are proportional to the electric
charge. This is in contrast with the $1/4$ supersymmetric GLPS
dyonic string that approaches the product of $AdS_3$ times a
squashed $3$-sphere with $1/2$ supersymmetry. In GLPS solution the
squashing is necessarily non-vanishing for non-vanishing gauge
coupling constant, while in our case the round $3$-sphere emerges
even in presence of nonvanishing gauge coupling.

One might naively expect that a double dimensional reduction of our
dyonic string might yield a novel black hole solution in $5D$ with
active hyperscalars. However,  we find that the resulting $5D$
metric has a naked singularity at the origin.

We conclude with mention of a selected open problems. The existence
of the supersymmetric dyonic string solution is encouraging with
regard to the string/M theory origin of the $6D$ model. The source
couplings we have found may provide additional information towards
that end. The existence of black dyonic strings in the $SU(2)_R$
gauged theory motivates a search for 'naturally' anomaly free such
models. We refer the reader to the introduction for what we mean by
'natural'. In any event, the string/M theory of origin of the matter
coupled $N=(1,0),\,6D$ gauged supergravities remains a challenging
open problem.

Here, we have begun to uncover some universal features of
supersymmetric solutions in which the sigma models play a nontrivial
role. For example, the emergence of tear-drop like metrics in the
space transverse to the brane. This is intimately related with
another potentially universal mechanism by which a submanifold of
the sigma model is identified with the transverse space. One
possible generalization might involve more intricate maps from the
transverse space to sigma model. It would be useful to find further
examples to establish whether the features found here continue to
persist in a larger class of supergravity models with supergravity
sectors.

\bigskip\bigskip

{\bf Acknowledgments}

The work of A.K. has been supported in part by the Turkish Academy
of Sciences via Young Investigator Award (TUBA-GEBIP), and the work
of D.C.J. and E.S. is supported in part by NSF Grant PHY-0314712,
and that of E.S in part by the Scientific and Technological Research
Council of Turkey (TUBITAK). E.S. would like to thank Feza
G\"{u}rsey Institute and Bo\~{g}azi\c{c}i University Physics
Department, where this work was done, for hospitality. We also thank
S. Deger and R. G\"uven for useful discussions.

\newpage

\begin{appendix}


\section{ Conventions }


We use the spacetime signature $(-+++++)$ and set
$\e^{+-ijkl}=\e^{ijkl}$. We define $\C_7=\C^{012345}$. The
supersymmetry parameter has the positive chirality: $\C_7\,\e=\e$.
Thus, $\C_{\mu\nu\rho}=
\ft16\,\e_{\mu\nu\rho\sigma\lambda\tau}\,\C^{\sigma\lambda\tau}\,\C_7$,
and for a self-dual 3-form we have
$S_{\mu\nu\rho}\C^{\mu\nu\rho}\e=0$.

The Hodge-dual of a $p$-form
\be F=\frac1{p!}\, dx^{\mu_1} \wedge \cdots dx^{\mu_p} F_{\mu_1\dots
\mu_p}\ ,\ee
is calculated using
\be *(dx^{\mu_1} \wedge \cdots dx^{\mu_p})=\frac1{(D-p)!}\,
\e_{\nu_1\dots\nu_{D-p}}{}^{\mu_1\dots\mu_p}\, dx^{\nu_1}\cdots
dx^{\nu_{D-p}}\ . \ee
The 't Hoof symbols are defined as
\be \rho^r_{\a\b}= {\rm tr}\,(\sigma_{\a}\, T^r\,{\bar\sigma}_{\b})\
,\quad\quad \eta^{r'}_{\a\b}= {\rm tr}\,(\bar \sigma_{\a}\,
T^{r'}\,{\sigma}_{\b})\ , \ee
where $ \sigma_{\a}=(1,-i \vec{\sigma} )$ are the constant van der
Wardeen symbols for $SO(4)$. These are real and  antisymmetric
matrices. It is easily verified that $\rho^r_{\a\b}$ is
anti-selfdual, while $\eta^{r'}_{\a\b}$ is selfdual. Their further
properties are
\bea
&& \rho^r_{\a\gamma}\, (\rho^s)^\gamma{}_\beta=
-\delta^{rs}\d_{\a\b}+\e^{rst}\,\rho^t_{\a\b}\ ,
\quad\quad\quad\quad\ \  {\rm idem} \ \eta^{r'}_{\a\b}\ ,\nn\w2
&& \rho^r_{\a\b} \rho^r_{\gamma\d} =
\delta_{\a\gamma}\d_{\b\d}-\delta_{\a\d}\d_{\b\gamma}
 -\e_{\a\b\gamma\d}\ ,\nn\w2
 &&\eta^{r'}_{\a\b} \eta^{r'}_{\gamma\d} =
\delta_{\a\gamma}\d_{\b\d}-\delta_{\a\d}\d_{\b\gamma}
 +\e_{\a\b\gamma\d}\ ,\nn\w2
 && \e^{trs}(\rho^r)_{\a\b}~(\rho^s)_{\gamma\d}=\d_{\b\gamma}~(\rho^t)_{\a\d}+{\rm 3\
 more}\ ,\quad\quad {\rm idem} \ \eta^{r'}_{\a\b}\ .
 \eea
For $SU(2)$ triplets, we use the notation:
\be X^{AB}= X^r\,T^r_{AB}\ ,\qquad X^r=\ft12 X^{AB}\,T^r_{AB}
 .\ee
%


\section{ The Gauged Maurer-Cartan Form and the $C$-Functions }


A convenient choice for the $Sp(n_H,1)/Sp(n_H)\times Sp(1)$ coset
representative $L$ is \cite{Gursey:1979tu}
\begin{equation}
L = \gamma^{-1} \left(\ba{ccc} 1 && t^\dagger\\
&\\ t  && \Lambda \ea \right)\label{cr}
\end{equation}
where $t$ is an $n_H$-component quaternionic vector $t^p\,\,
(p=1,...,n_H)$, and
\be \gamma= (1-t^\dagger\, t)^{1/2}\ ,\qquad \Lambda=
\gamma\,(I-t\, t^\dagger)^{-1/2} \ . \ee
Here, $I$ is an $n_H\times n_H$ unit matrix, and $\dagger$ refers to
quaternionic conjugation, and it can be verified that $\Lambda t=t$.
The gauged Maurer-Cartan form is defined as
\begin{equation}
L^{-1} D_\mu L = \left(\ba{ccc} Q_\m && P_\m^\dagger \\&\\ P_\m && Q'_\m
\ea \right)\ ,
\end{equation}
where $D_\mu L$ is given in \eq{mc3}, with $T^r$ representing three
anti-hermitian quaternions (in the matrix representation of
quaternions $T^r=-i\,\s^r/2$)  obeying \be [T^r,T^s]=\e^{rst}T^t \ee
and $T^{I'}$ represents a subset of $n_H\times n_H$ quaternion
valued anti-hermitian matrices spanning the algebra of the subgroup
$K\subset Sp(n_H)$ that is being gauged. A direct computation gives
\begin{eqnarray}
Q_\m &=& \frac12\, \gamma^{-2}\,\left(D_\m t^\dagger t - t^\dagger
D_\m t\right)  -A_\mu^rT^r\la{aq1}\w2
Q'_\m &=& \gamma^{-2}\,\left(-t D_\mu t^\dagger +\Lambda
D_\mu\Lambda +\ft12 \partial_\mu (t^\dagger t) I \right) -A_\mu^{I'}
T^{I'}\ ,\la{aq2}\w2
P_\m &=& \gamma^{-2} \Lambda D_\m t \ ,\la{aq3}
\end{eqnarray}

where
\be D_\mu t=\partial_\mu t +t\, T^r A_\mu^r -A_\mu^{I'}
T^{I'}\,t\ . \la{ddf}\ee

The $C$ functions are easily computed to yield
\bea C^r&=&L^{-1}T^r L=\gamma^{-2}\left(
                        \begin{array}{ccc}
                          T^r && T^r t^\dagger \\&\\
                          -t T^r && -t T^r t^\dagger\\
                        \end{array}
                      \right)\la{crr}
                      \w4
C^{I'}&=& L^{-1}T^{I'} L = \gamma^{-2}\left(
                             \begin{array}{ccc}
                               -t ^\dagger T^{I'} t && -t^\dagger T^{I'}\Lambda \\&\\
                               \Lambda T^{I'}t && \Lambda T^{I'}\Lambda \\
                             \end{array}
                           \right)\la{ci}
\eea
%


\section{The Model for $Sp(1,1)/Sp(1)\times Sp(1)_R$}


This coset, which is equivalent to $SO(4,1)/SO(4)$, represents a
4-hyperboloid $H_4$. In this case we have a single quaternion $
t=\phi^\a\,\sigma_\a$, and the vielbein becomes
\be V_\a^{A'A}= \gamma^{-2}\,\sigma_\a^{A'A}\ .\ee
It follows from the definitions \eq{vg} and \eq{j} that
\be g_{\a\b}={2\over (1-\phi^2)^2}\,\delta_{\a\b}\ ,\quad\quad
J^r_{\a\b}= {2\,\rho^r_{\a\b}\over (1-\phi^2)^2}\ .\label{jr} \ee
We also introduce a basis in the tangent space of $H_4$
\be
V_\a{}^{\ua}=\fr{\sqrt{2}}{1-\f^2}\,\, \d^{\ua}_{\a}\, \,
.\label{av} \ee
The $Sp(1)_R$ connection $Q_\m^r$ can be found from \eq{aq1} as
\be Q_\mu^r=-2\,{\rm tr}\,(Q_\m T^r)\,= {1\over 1-\phi^2}\,\left(2 \rho^r_{\a\b}
\partial_\mu\phi^\a\,\phi^\b-A_\mu^r\right)\ .\label{qsp1}
\ee
With the above results at hand, the Lagrangian can be written as
\bea
e^{-1}{\cal L} &=& R\,- \ft14 (\p\vf)^2- \ft12 e^\vf\,
G_{\mu\nu\rho}G^{\mu\nu\rho}- \ft14\,e^{\ft12\vf}\,
F^r_{\m\nu}\,F^{r \mu\nu}\,- \ft14\,e^{\ft12\vf}\,
F^{r'}_{\m\nu}\,F^{r'\mu\nu}\,\nn\w2
&&-{4\over (1-\phi^2)^2}\,D_\mu\phi^\a D^\mu\phi^\b\,\d_{\a\b}
- {6 e^{-\ft12\vf}\over (1-\phi^2)^2 }\,\left[ g_R^2 +g'^2
(\phi^2)^2\right]\ , \eea
where the covariant derivatives are defined as
\be D_\m\phi^{\a}= \partial_\mu \phi^{\a}-\ft12 g_R A_\mu^r
(\rho^r)^{\a}{}_{\b}\,\,\phi^{\b} - \ft12 g' A_\mu^{r'}
(\eta^{r'})^{\a}{}_{\b}\,\,\phi^{\b} , \ee
and we have re-introduced the gauge coupling constants $g_R$ and
$g'$. The supersymmetry transformation rules are
\bea
\d \psi_\m &=& D_\m \ve + \ft1{48} e^{\ft12\vf}
G_{\nu\s\rho}^+\,\C^{\nu\s\rho}\, \C_\m\,\ve \ ,\w2
\d\chi &=&\ft14\left( \C^\m\p_\m \vf  -\ft16 e^{\ft12\vf}
G_{\m\nu\rho}^-\,\C^{\m\nu\rho} \right)\ve\ , \w2
\d \l_A^r &=& -\ft18 F_{\m\nu}^r\C^{\m\nu}\ve_A
    - g_R{e^{-\ft12\vf}\over 1-\phi^2}\,T^r_{AB}~\ve^B \ ,
\w2 %
\d \l_A^{r'} &=& -\ft18 F_{\m\nu}^{r'}\C^{\m\nu}\ve_A
    + g'e^{-\ft12\vf} {\phi^\a\phi^\b \over 1-\phi^2}\, ({\bar\s}_\a T^{r'}\s_\b)_{AB}~\ve^B \
    ,\w2
\d\psi^{A'} &=&  {1\over 1-\phi^2}\, D_\mu
\phi^\a\,\s_a^{A'A}\,\ve_A \ , \eea

where $D_\m\ve_A = \nabla_\m\ve_A + Q_\m^r (T^r)_ A{}^B \ve_B$, with
$\nabla_\mu$ containing the standard torsion-free Lorentz connection
only, and $Q^r$ is defined in \eq{qsp1}.

\end{appendix}


\end{document}